\documentclass[aps,pre,twocolumn,showpacs]{revtex4}

\usepackage{graphicx}
\usepackage{amsmath}
\usepackage{amssymb}
\usepackage{bm}

\begin{document}
%_______________
%
%

\title{Finite-size anisotropy in  statistically uniform porous media}

\author{Zbigniew Koza and Maciej Matyka}
\affiliation{Institute of Theoretical Physics,University of Wroc{\l}aw,
        pl.\ M.\ Borna 9, 50-204 Wroc{\l}aw, Poland}

%\author{Maciej Matyka}

%\affiliation{Institute of Theoretical Physics, University of Wroc{\l}aw, pl.\
%M.\ Borna 9, 50-204 Wroc{\l}aw, Poland}

\author{Arzhang Khalili}
\affiliation{Max Planck Institute for Marine Microbiology, Celsiusstrasse 1,
D-28359 Bremen, Germany }
\affiliation{Jacobs University Bremen, Campus Ring 1, D-28759 Bremen, Germany }

\date{\today}

%_______________
%
\begin{abstract}
%_______________
%
%
%
%
%
Anisotropy of the permeability tensor in statistically uniform porous media of
sizes used in typical computer simulations is studied. Although such systems
are assumed to be isotropic by default, we show that de facto their anisotropic
permeability can give rise to significant changes of transport parameters such
as permeability and tortuosity. The main parameter controlling the anisotropy
is $a/L$,  being the ratio of the obstacle to system size.  Distribution of the
angle $\alpha$ between the external force and the volumetric fluid stream is
found to be approximately normal, and the standard deviation of $\alpha$ is
found to decay with the system size as $(a/L)^{d/2}$, where $d$ is the space
dimensionality. These properties can be used to estimate both
anisotropy-related statistical errors in large-scale simulations and the size
of the representative elementary volume.
\end{abstract}

% 47.15.G- == Low-Reynolds-number (creeping) flows
% 47.56.+r == Flows through porous media
% 91.60.Np == Permeability and porosity


\maketitle

%_____________________
%
\section{Introduction}
%_____________________

A standard method of modeling a uniform, isotropic porous medium (e.g.\ a
column of sand) is to place randomly many identical objects that are
impermeable to fluid (e.g.\ solid spheres) in an initially empty volume
\cite{Martys92,Koponen97,Nabovati07,Verberg99,Kostek92,Matyka08,Basagaoglu06,Zhang95}.
Since the objects are placed uniformly in the whole system, one might expect
that randomness in their exact locations is irrelevant in the sense that the
bulk volumetric fluid stream will be parallel to the external force (e.g.\
gravitation). This would be the case if the system was large enough. However,
in computer simulations and in artificial laboratory systems (used in particle
image velocimetry measurements \cite{Zerai05}), usually relatively small
systems are utilized that contain at most a few thousands of "grains"---far
less than billions of sand grains in a typical experimental setup. Since
randomly distributed grains tend to form channels of random orientations, small
porous systems
%!!!!, especially those with small porosity,
are very sensitive to
local fluctuations of the grain distribution. Under such conditions the
direction of the volumetric fluid stream can differ significantly from that of
the external force. Consequently, a system that  was supposed to be isotropic,
may de facto be rather highly anisotropic. The aim of this paper is a detailed
analysis of this phenomenon in a two-dimensional (2D) flow.

A porous medium is anisotropic to flow if the permeability tensor is
anisotropic. Usually anisotropy of  the permeability tensor is either assumed
explicitly \cite{Bear72} or it appears naturally as an expected consequence of
a microscopic model \cite{Scholes07,Boudreau06,Alam06,Ohkubo08}. In the former
case one works entirely on a macroscopic level, whereas the latter approach
tries to connect the observed macroscopic anisotropy of real porous materials
with their microscopic geometry and structure. Permeability anisotropy caused
by a finite size of a model system has not been regarded as an important factor
so far, although some research techniques, e.g.\ numerical simulations,
concentrate on artificially small porous systems. The reason for this lies in
the fact that  numerical flow simulations in complex porous geometries are
extremely tedious and require extensive computer power and resources. A common
strategy has been to perform calculations for just a few systems that are as
large as possible \cite{Cancelliere90,Basagaoglu06}. In contrast to this, here
we solve the flow equations for hundreds or even thousands of different porous
systems of small to medium sizes and then extrapolate the results to the limit
of an infinitely large system.
This method was already used in \cite{Matyka08} to detect a small, systematic
deviation of the flow tortuosity from several theoretical formulas, with an
\emph{ad hoc} interpretation of this phenomenon as a consequence of the
finite-size anisotropy. Therefore, in this paper we present a systematic study
of finite-size anisotropy in a two-dimensional model of statistically uniform
porous media.

The structure of the paper is as follows. Section \ref{sec:model} specifies the
model and the numerical techniques  used.  Main results are provided in Sec.\
\ref{sec:results}. Next, in Section \ref{sec:4} we develop a simple theory to
account for the asymptotic behavior of the angle between the external force and
the volumetric fluid flux. Finally, the results are discussed in Sec.\
\ref{sec:discussion}.

% ************************************
% ************ MODEL *****************
% ************************************

\section{Model    \label{sec:model}}

In this study we use a model of freely overlapping squares
\cite{Koponen96,Koponen97,Matyka08}. In essence, this is a two-dimensional
lattice model with a porous matrix modelled as a union of freely overlapping
identical solid squares of size $a\times a$ lattice units (l.u.) placed
uniformly at random locations in a square lattice $L\times L$ l.u.\ ($1 \le a
\ll L$). The squares are fixed in space but free to overlap, and their sides
coincide with the underlying lattice. The remaining void space is filled with a
fluid. A constant, external force is imposed on the fluid to model the gravity
and we allow an angle ($\beta$) between the force and the system side to be
arbitrary (note that in \cite{Koponen96,Koponen97,Matyka08} only the case
$\beta=0$ was considered).  Periodic boundary conditions are imposed in both
directions to minimize finite-size effects. The porosity ($\phi$) is calculated
as the ratio of unoccupied lattice nodes to the entire system volume ($L^2$). The
flow equations are solved in the creeping flow regime using the Lattice
Boltzmann Model (LBM) \cite{Succi01} with a single relaxation time collision
operator \cite{Bhatnagar54} (see \cite{Matyka08} for implementation details).

The model has three adjustable parameters: $\phi$, $a$, and $L$. The first one
corresponds directly to the macroscopic porosity. The value of $a$ affects the
percolation threshold $\phi_c$, which is a decreasing function of $a$ from
$\phi_c\approx 0.4073$ (the standard site percolation threshold, $a=1$)
\cite{Sahimi93} to $\phi_c \approx 0.3333$ (the continuous percolation
threshold of aligned squares, $a\to\infty$) \cite{BakerStanley04}. As the model
is solved using the LBM method without a numerical grid refinement
\cite{Matyka08}, the minimum value of $a$ is 4 (this is the minimum length
scale for the LBM method to resolve the macroscopic Navier-Stokes equations
\cite{Succi01}). The value of $L$ controls the finite-size effects through the
dimensionless ratio $a/L$, which should be as small as possible to mimic an
infinite system.

Anisotropy of fluid flow in the above-defined model will be investigated
through Darcy's law \cite{Bear72}
\begin{equation}
  \label{eq:Darcy}
   \bm{q}= \bm{K}\hat{\bm{g}},
\end{equation}
where ${\bm q}$ is the volumetric fluid flux, $\bm{K}$ is a symmetric tensor of
the hydraulic conductivity, and $\hat{\bm{g}}$ is the unit vector in the
direction of the gravitational field.

%In practice, the system size $L$ is bounded from above by available computing power.

% Following \cite{Koponen96,Koponen97}, in all simulations the parameter $a$ has
% been set to 10.

% **************************************
% ************ RESULTS *****************
% **************************************

\section{Numerical results    \label{sec:results}}

\subsection{Tests on $\bm{K}$}

The basic concept of transition from microscopic laws of
hydrodynamics to macroscopic laws of transport in porous media is the
representative elementary volume (REV), i.e.\ the smallest volume such that a measurement over it will
yield a value representative of the whole \cite{Bear72}. Darcy's  law
(\ref{eq:Darcy}) is, in principle, applicable only to systems that are larger
than an REV, whereas significant anisotropy is expected in
systems smaller
than an REV.
Hence, the primary question is whether or not Eq.~(\ref{eq:Darcy}) can be
used to study anisotropy in small-size systems. To answer this we performed
several simulations on $\bm{K}$,  with its elements computed from
$\bm{q(\hat{g})}$ for $\bm{\hat{g}} = \bm{\hat{x}}$ and $\bm{\hat{g}} =
\bm{\hat{y}}$.

First, the symmetry of  $\bm{K}$ was examined by quantifying the value of a
dimensionless parameter given by
\begin{equation}
  \epsilon = \frac{|K_{xy} - K_{yx}|}{K_{xx} + K_{yy}}.
\end{equation}
Furthermore, by choosing  $L=100$ l.u., $a=4$ l.u., and $\phi=0.7,0.9$,
eighty  different (i.e., fourty systems for each $\phi$) statistically uniform
porous systems were constructed, for which $\epsilon <$ 0.5\% was found. This
ensures that $\bm{K}$ is symmetric within  $0.5\%$ numerical errors in its
elements. In a subsequent analysis  we enforced $\bm{K}$-symmetry via replacing
its off-diagonal elements ($K_{xy}$ and $K_{yx}$) by their arithmetic mean,
which ensures that $\bm{K}$ is diagonizable.

\begin{figure}
\includegraphics[width=0.45\columnwidth]{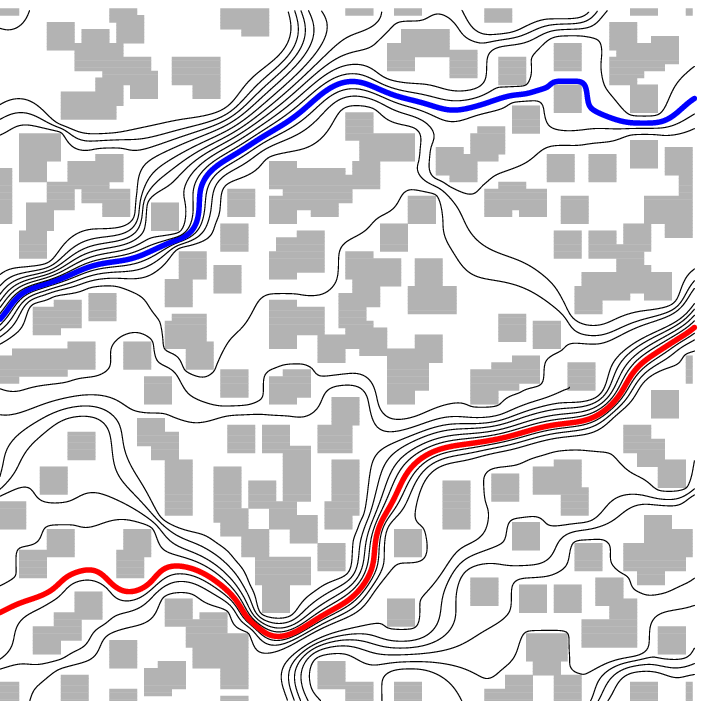}\hspace{2ex}
\includegraphics[width=0.45\columnwidth]{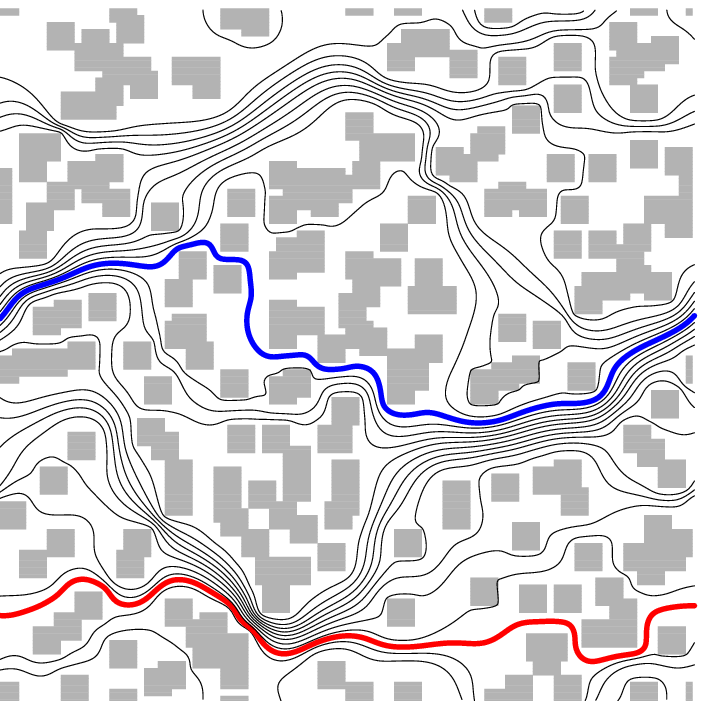}
\caption{Streamlines through the same porous system
    ($L=100$ l.u., $a=4$ l.u., $\phi=0.7$) for two different alignments of the
    external force $\bm{g}$. The grey squares represent the solid part of the
    medium, and the remaining space is open to fluid flow.
    Left panel: $\bm{\hat{g}}$ is parallel to the $x$ axis ($\beta=0$) and the
    specific discharge $\bm{q}$ makes an angle $\alpha\approx 21^{\circ}$ with
$\bm{\hat{x}}$.
    Right panel:
    $\beta\approx-22^{\circ}$ (calculated from Eq.~(\protect\ref{eq:angle})) and
    the  angle between the specific discharge $\bm{q}$ and the $x$ axis is
    $\alpha\approx 0.7^{\circ}$.
    For the ease of display, two selected streamlines and their counterparts
    in both panels are given in color.} \label{fig:1abc}
\end{figure}

Second, the tensorial properties of $\bm{K}$ were examined by checking whether
Eq.~(\ref{eq:Darcy}) can be used for an arbitrary $\bm{\hat{g}}$.
In particular, this equation predicts that if the mean flow direction
($\bm{q}$) is aligned with the $x$-axis, the angle $\beta$ between the
external force ($\bm{{g}}$) and the $x$-axis should satisfy
\begin{equation}
 \label{eq:angle}
 \tan\beta = -\frac{K_{xy}}{K_{yy}}.
\end{equation}
This relation was examined for several systems, of which one is shown in
Fig.~\ref{fig:1abc}, wherein, two streamline patterns for the  same system
($L=100$ l.u., $a=4$ l.u., $\phi=0.7$) with different $\bm{g}$-orientations are
visualized.
In the left panel, the external force is parallel to the (horizontal) $x$-axis
($\bm{\hat{g}}= \bm{\hat{x}}$), resulting in an angle
of $\alpha \approx 21^\circ$ between the vector
of the specific discharge ($\bm{q}$) and the $x$ axis.

In the right panel, a force of the same magnitude makes an angle $\beta \approx
-22^\circ$ computed from (\ref{eq:angle}); as expected, the angle between
$\bm{\hat{q}}$ and the $x$ axis practically vanishes ($\alpha \approx -0.7^\circ$).

%_______________________________ FIG-1ab
%________________________________________

%______________________________

\subsection{Tests on $\alpha$}
%______________________________

\begin{figure}
\includegraphics[width=0.85\columnwidth]{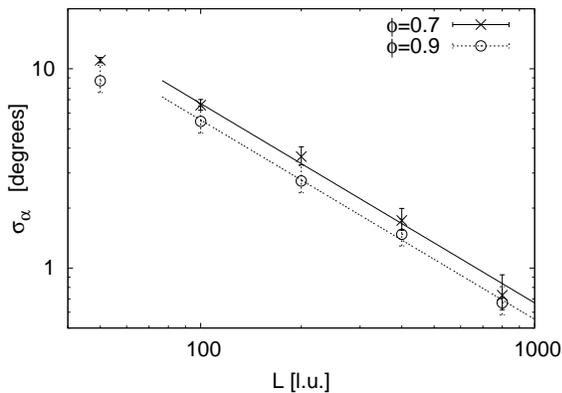}
\caption{$\sigma_\alpha = \sqrt{\langle \alpha^2 \rangle}$
 as a function of $L$ for $a=4$ l.u.\ and two porosities $\phi=0.7, 0.9$, with
 error bars at  95\% confidence level. The lines represent fits to the power
 law $\sigma_\alpha \propto L^{-1}$ for $L\ge 100$ l.u. \label{fig:2}}
\end{figure}

A natural  measure of anisotropy for a particular porous system is the angle
between the vectors $\bm{\hat{g}}$ and $\bm{q}$. As this angle depends on the
orientation of $\bm{\hat{g}}$, following standard procedures in computer
simulations, we fix $\bm{\hat{g}} =\bm{\hat{x}}$. We verified that in this case
the numerical value of $\alpha$ (angle between $\bm{q}$ and $\bm{\hat{g}}$)
satisfies $\langle\alpha\rangle \approx 0$, which follows from symmetry
arguments, and then calculated
\begin{equation}
 \label{eq:sigma-alpha}
 \sigma_\alpha = \sqrt{\langle \alpha^2 \rangle}.
\end{equation}
In the above equation, $\langle\cdots\rangle$ denotes an average over different
random porous systems. The results for $a=4$ l.u., $\phi =0.7, 0.9$, and
several system lengths $L$ are shown in Fig.~\ref{fig:2}. For $L \ge 100$ l.u.\
the data were fitted to
\begin{equation}
 \label{eq:delta}
  \sigma_\alpha \propto L^{-\delta},
\end{equation}
which yielded $\delta\approx 0.96(6)$ for $\phi=0.7$ and $\delta\approx 1.00(3)$
for $\phi=0.9$. This suggests $\delta=1$, i.e.
\begin{equation}
 \label{eq:delta=1}
  \sigma_\alpha \propto L^{-1}, \quad L \gg 1.
\end{equation}
This relation does not hold for small $L$ ($L\lesssim 50$ in Fig.~\ref{fig:2}),
for some realizations of such systems are likely to exhibit extreme anisotropy
with $\alpha$ so large that $\sin\alpha$ cannot be approximated by $\alpha$
(for $L=50$ the angle between $\bm{q}$ and $\bm{\hat{g}}$ can be as large as
$45^\circ$).

%____________________________________ FIG-2
%______________________________________

\begin{figure}
\includegraphics[width=0.85\columnwidth]{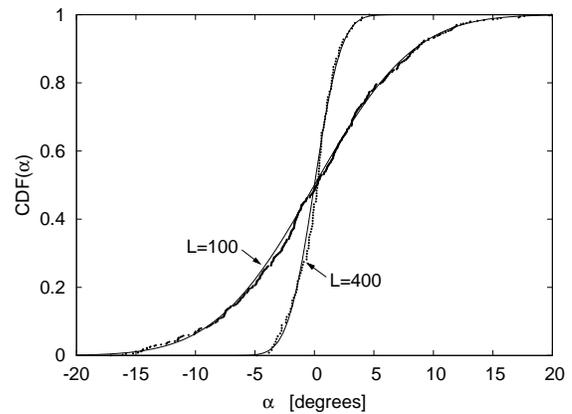}
\caption{Cumulative distribution function (CDF) of $\alpha$ for $\phi=0.7$,
$a=4$ l.u.\ and two system sizes $L=100$ and $L=400$ l.u.\
    (dots). The empirical CDF was determined  using 500
(for $L=100$) and 80 ($L=400$)
numerical samples. Solid lines represent theoretical CDF of
the normal distribution, Eq.~(\ref{eq:F}), with
$\sigma_\alpha=\sqrt{\langle \alpha^2 \rangle}$.
 \label{fig:3}}
\end{figure}

Next we investigated statistical distribution  of $\alpha$-values in different
random systems with fixed $L$, $a$, and $\phi$. In all cases this
distribution closely resembled the normal distribution $N(0, \sigma_\alpha^2)$.
Qualitative verification of this conjecture is presented in Fig.~\ref{fig:3},
which depicts the empirical cumulative distribution function (CDF) for two
different system sizes $L$ (small symbols) together with the corresponding
theoretical CDFs of the normal distribution,
\begin{equation}
  \label{eq:F}
   F(\alpha) = \frac{1+\mathrm{erf}\,(\alpha/\sqrt{2}\sigma_\alpha)}{2}.
\end{equation}
The numerical data are in good agreement with (\ref{eq:F}). A quantitative
comparison of the $\alpha$-distribution with $N(0,\sigma_\alpha^2)$ was performed
using  the Kolmogorov-Smirnov test (at confidence level 95\%). Out of all data
points shown in Fig.~\ref{fig:2} only the one corresponding to $L=50$ and
$\phi=0.7$ did not pass the test, which in part is due to extremely large
number of different samples (2000) used.

%___________________________________ FIG-3
%
%___________________________________
%

\begin{figure}
\includegraphics[width=0.9\columnwidth]{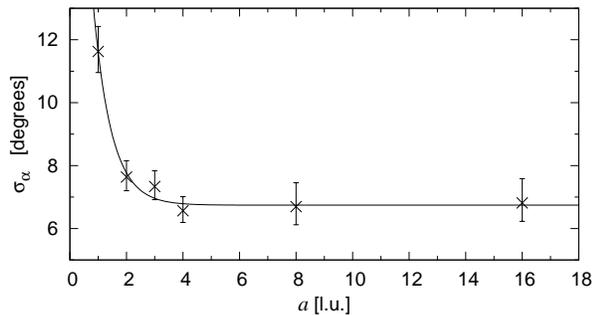}
\caption{$\sigma_\alpha$ as a function of $a$ for  $\phi=0.7$
 and $a/L= 0.04$ ($\times$~symbols). The data come from $N=500$ independent
 porous systems for $a\le4$ l.u.\ and $N=200$ for $a>4$ l.u. The error bars
 were calculated at the 95\% confidence level. The solid line represents the best fit to
 Eq.~(\protect\ref{eq:fit-exp}). \label{fig:4}}
\end{figure}

As  mentioned before, the value of  $a$ determines the percolation threshold
$\phi_c$ for small porosities, and can be considered as a relevant parameter
independent of $\phi$ and $L$.
For porosities  much larger than $\phi_c$, however, the connectedness and
overlapping of individual randomly generated solid squares  becomes irrelevant.
In this case, using scaling arguments, one can expect that $\phi$ and $a/L$ are
the only relevant parameters. Mathematically, this can be formulated as a
similarity ansatz:
\begin{equation}
 \label{eq:similarity}
   \sigma_\alpha(a,L,\phi) \approx \Psi(a/L, \phi),\quad \phi\gg\phi_c, L\gg a.
\end{equation}
where $\Psi$ is a similarity function. A direct numerical verification of this
conjecture is difficult, as it requires averaging over many independent
samples, which is rather a time-consuming job for large $a$. Instead of this,
we concentrated on a single parameter set with $\phi=0.7$ and $a/L=0.04$ that
led to results demonstrated in Fig.~\ref{fig:4}
%_________________________________ FIG-4
%
%
%___________________________________
shown as cross symbols. These data were fitted to an \emph{ad-hoc} formula
\begin{equation}
    \label{eq:fit-exp}
    \sigma_\alpha(a) = c_1 + c_2\exp(-a/c_3)
\end{equation}
with three adjustable parameters $c_1$, $c_2$, and $c_3$. The best-fit value of
$c_3\approx 0.6$ indicates that the approximation (\ref{eq:similarity}) can be
safely used for $a\gtrsim 4$.

\begin{figure}
\includegraphics[width=0.9\columnwidth]{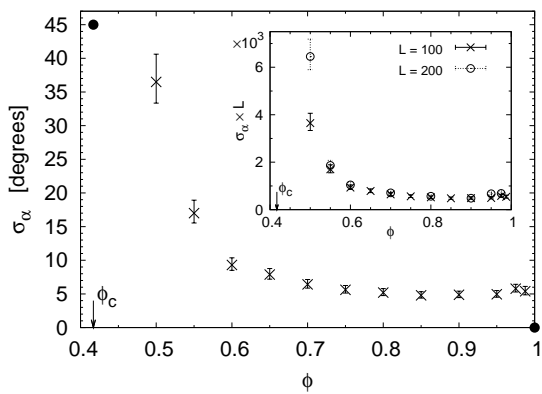}
\caption{$\sigma_\alpha $ (in degrees) as a function of porosity $\phi$
 for $L=100$ l.u. Inset: the product $\sigma_\alpha L$ (deg.\
 $\times$ l.u.) for $L=100$ l.u.\ (cross symbols) and $L=200$ l.u.\ (circle symbols). All data
 obtained for $a=4$ l.u.; error bars at 95\% confidence level obtained from 200
independent porous systems. The arrows show the percolation threshold
$\phi_c\approx 0.417$. Filled circles show theoretical values of
 $\sigma_\alpha$ for $\phi=\phi_c$, $L\to\infty$ ($45^\circ$) and $\phi=1$
 ($0^\circ$). \label{fig:5}}
\end{figure}

Finally, we investigated the dependence of $\sigma_\alpha$ on porosity. One
expects that $\sigma_\alpha$  should decrease from $\approx 45^\circ$ at the
percolation threshold $\phi_c$ (a single, randomly oriented conducting channel)
to $0^\circ$ at $\phi=1$ (completely permeable system).
As shown in Fig.~\ref{fig:5}, our numerical results generally agreed with this
picture. However, $\sigma_\alpha$ did not converge to its limiting value $0$ as
$\phi\to 1$. Instaed, it  saturated at a positive value, which is independent
of $\phi$.
Due to this rather unexpected result, we ensured that neither discretization
errors nor large relaxation times affect the numerical data obtained for
large porosities. We also verified that the system size used, $L=100$, is
sufficiently large for relation (\ref{eq:delta=1}) to hold.
This is clearly seen in the inset of Fig.~\ref{fig:5}, which depicts the
product $\sigma_\alpha L$ for $L=100$ and $L=200$. The data for different $L$
collapsed in a broad range of $\phi \gtrsim 0.55$. Porosities less than
$\approx0.55$ are in a vicinity of the percolation critical point, at which
$\sigma_\alpha$ is expected to converge to $45^\circ$ as $L\to\infty$, and
hence the product $\sigma_\alpha L$ should diverge at $\phi_c$. As the system
size $L\to\infty$, it is possible that the the porosity range, for which
scaling relation (\ref{eq:delta=1}) does not hold, diminishes according to a
power law. This behavior is  a typical finite-size effect near a critical point
\cite{Hunt05}.

%_________________________________ FIG-5
%
%______________________________________

\begin{figure}
\begin{center}
\includegraphics[width=0.45\columnwidth]{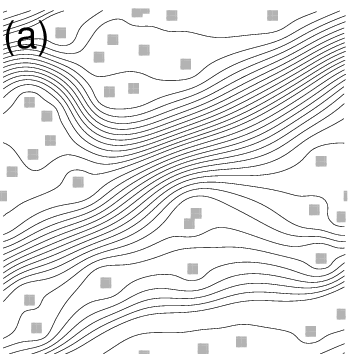}\hspace{2ex}
\includegraphics[width=0.45\columnwidth]{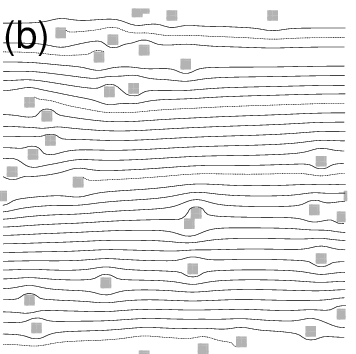}
\end{center}
\caption{Streamlines in a high-porosity system
 ($\phi=0.95$) with $L=100$ l.u. and $a=4$ l.u. for two different flow types:
 (a)~hydrodynamic ($\alpha \approx 15^\circ$); (b)~electric ($\alpha \approx
 0.6^\circ$). The electric flow was calculated from a solution of the Laplace equation with
 periodic boundary conditions and a constant electric field parallel to the $x$-axis.
 Note that the distribution of obstacles and
 the orientation of an external body force is identical in both panels.   \label{fig:6ab}}
\end{figure}

To explore the reason why $\sigma_\alpha$ does not tend smoothly to 0 as $\phi$
approaches 1, we inspected the streamlines in high-porosity systems exhibiting
large anisotropy. An extreme example of such a system, generated with
$\phi=0.95$, is shown in Fig.\ \ref{fig:6ab}a. At this high porosity,
overlapping of individual obstacles is negligible, and the solid part of the
system is made up of separate islands (that could correspond, for example, to a
cross-section of a porous medium made of parallel fibers
\cite{Koponen98,Araujo06}). Because the obstacles were placed uniformly and
\emph{randomly} in the whole system, their local concentration varies, and they
form several larger groups of obstacles with relatively small distances between
group members. Since fluid flux through a 2D channel is proportional to its
width \emph{squared}, most of the fluid flow takes place in relatively wide
`channels' between the groups. In other words, owing to no-slip boundary
conditions on the obstacle surfaces, the fluid passes most easily in the
inter-connected regions of low local obstacle concentration (high local
porosity), whereas the regions of high local obstacle concentration (low local
porosity)---even if occupied by separate obstacles---act effectively as large,
almost impenetrable barriers. This solid-fluid `repulsion' effect is not
present in electric current flows (for the current intensity is proportional to
the \emph{first} power of a conductive channel width). For this reason,  a
high-porosity system which is highly anisotropic to fluid flow ($\alpha \approx
15^\circ$ in Fig.\ \ref{fig:6ab}a) exhibits a marginal anisotropy to electric
current flow ($|\alpha| < 1^\circ$) as depicted  in Fig.\ \ref{fig:6ab}b.

%_________________________________ FIG 6ab
%
%
%__________________________________

%_____________________________________
%
\subsection{Tests on principal values}
%_____________________________________

Mathematically, a porous system is anisotropic to flow if and only if at least
two of the principal values of $\bm{K}$ are diffrent. In the present case
$\bm{K}$ has two eigenvalues (principal permeabilities) $K^+$ and $K^-$ which
can be ordered such  that $K^+ \ge K^-$. Their ratio,
\begin{equation}
 \label{eq:r}
   0 \le r = \frac{K^-}{K^+} \le 1,
\end{equation}
is equivalent to the ratio of the  minimum to maximum permeabilities of a given
porous system, and hence is a proper measure of its anisotropy
\cite{Scholes07}. The more $r$ deviates from $1$, the more anisotropic the
system is.

\begin{figure}
\includegraphics[width=0.9\columnwidth]{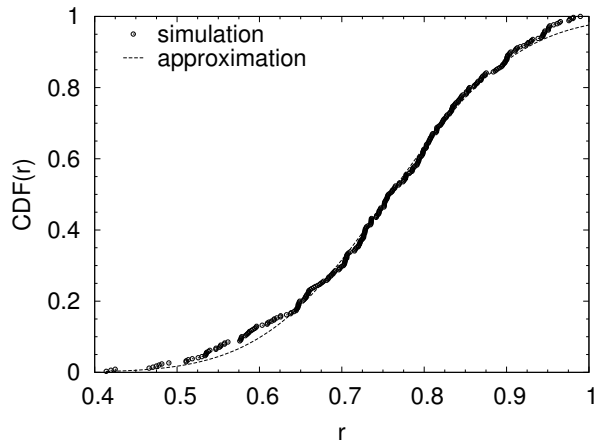}
\caption{Cumulative distribution function of
    $r=K^-/K^+$   for
    $\phi=0.7$, $a=4$ l.u., $L=100$ l.u., calculated using
    $N=340$ different porous
    systems.
    Dashed line represents the best fit to the CDF of the normal
    distribution with $\langle r \rangle \approx  0.75$
    and $\sigma_r \approx 0.12$.
 \label{fig:7}}
\end{figure}

We first verified that, as expected, the angle between the main principal
axis and the $x$-axis was distributed uniformly in the range $(-\pi/2,\pi/2]$
(data not shown). Then the CDF of $r$ was determined for a particular case with
$\phi=0.7$, $a=4$ l.u., and $L=100$ l.u.\
As can be seen in Figure~\ref{fig:7},
the distribution of $r$ can be quite well fitted to the normal distribution $N(0.75, 0.12^2)$.
However, this is only an approximation, as in the present case $\mathrm{CDF}(r)\equiv1$
for $r \ge 1$.

%
%______________________________ FIG-7
%
%
%______________________________

%%%%%%%%%%%%%%%%%%%%%%%%%%%%%%%%%%%%%
%%%%       SECTION IV            %%%%
%%%%%%%%%%%%%%%%%%%%%%%%%%%%%%%%%%%%%

%_____________________________________________________________
%
\section{Distribution of $\alpha$ for large $L$ \label{sec:4}}
%_____________________________________________________________

Consider a porous system of size $L \times L$ l.u.\ subject to an external
force along the $x$ axis. Let $\Delta \bm{r}$ denote the
total displacement of a fluid particle  as it passes the system between the
opposite boundaries.
%Let $\Delta x$ and $\Delta y$ denote the $x$ and $y$ components
%of $\Delta \bm{r}$, respectively.
While the $x$-component of $\Delta \bm{r}$ is a constant (equal to the system
size $L$), the $y$ component (which we shall call lateral displacement and
denote $\Delta y$) varies for different streamlines. If we calculate the
average $\langle \Delta y \rangle$ over all fluid particles, the angle $\alpha$
between the volumetric fluid flux $\bm{q}$ and the $x$ axis will satisfy
\begin{equation}
  \label{eq:tan_alpha}
    \tan \alpha = \frac{\langle \Delta y\rangle}{L}.
\end{equation}
If $\alpha$ is sufficiently small, this equation simplifies to
\begin{equation}
  \label{eq:alpha_approx}
    \alpha \approx \frac{\langle \Delta y\rangle}{L}.
\end{equation}

\begin{figure}
\includegraphics[width=0.3\columnwidth]{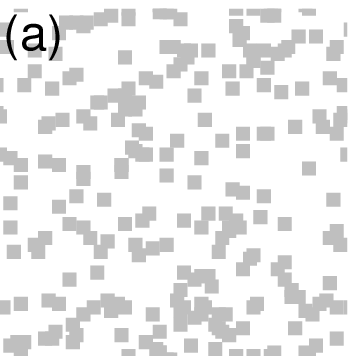} $\;\;$
\includegraphics[width=0.3\columnwidth]{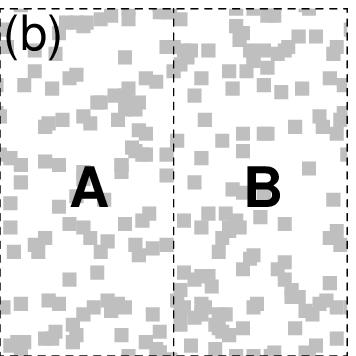} $\;\;$
\includegraphics[width=0.3\columnwidth]{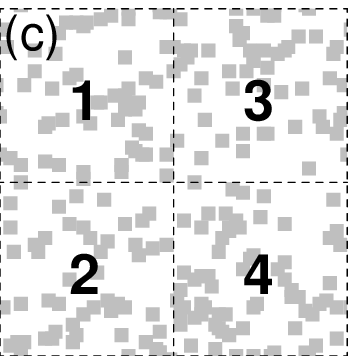}
\caption{A two-dimensional porous system (a), which can be regarded as a
 two-layer system perpendicular to the external
 force  (b) or a group of four smaller subsystems (c).
 \label{fig:8}}
\end{figure}

Let us consider a porous system of size $2L \times 2L$ l.u.\ and porosity
$\phi$. As shown in Fig.~\ref{fig:8}, it can also be regarded as two subsystems
of size $2L\times L$ l.u.\ (labelled $A$, $B$) or four  subsystems of size
$L\times L$ l.u.\ (labelled $1,2,3,4$).
Each of the small subsystems has its own permeability tensor $\bm{K}_j$,
volumetric fluid flux $\bm{q}_j$, angle $\alpha_j$ between the $x$ axis and
$\bm{q}_j$, and mean lateral displacement $\langle\Delta y_j\rangle$ with
$j=1,\ldots,4$. Since the distribution of obstacles is uniform, porosities of
each small subsystem is approximately equal to $\phi$, and the mean lateral
displacements $\langle\Delta y_j\rangle$ can be considered as independent random variables
drawn from the same distribution. Subsystems 1 and 2 form  layer $A$ orthogonal
to the external force. One may expect that the fluid streams passing
through subsystems 1 and 2 are approximately the same in magnitude, and so the
mean lateral displacement of the fluid, as it passes through layer $A$, can be
approximated by
\begin{equation}
  \langle\Delta y_A\rangle
     \approx
  \frac{\langle\Delta y_1\rangle + \langle\Delta y_2\rangle}{2}
\end{equation}
%%

%________________________________ FIG-8
%
%
%______________________________________

Similarly, the mean lateral displacement in the layer $B$ can be approximated
by $\langle\Delta y_B\rangle \approx (\langle\Delta y_3\rangle + \langle\Delta
y_4\rangle)/2$. The mean lateral displacements of the fluid in layers $A$ and
$B$ are practically independent of each other. This can be justified by an
example of soil made of several horizontal and anisotropic layers---in this
case the mean flow direction in a layer will depend only on the permeability
tensor of this layer. This implies that the total lateral displacement of the
fluid in the whole system ($\langle \Delta y \rangle $) is approximately given
by
\begin{equation}
  \label{eq:2L}
  \langle \Delta y \rangle
    \approx
  \langle\Delta y_A\rangle + \langle\Delta y_B\rangle
    \approx
  \frac 12 \sum_{j=1}^{4}\langle\Delta y_j\rangle.
\end{equation}

If $L$ is large, then $\alpha$ becomes sufficiently small for approximation
(\ref{eq:alpha_approx}) to be valid. In this case Eqs.~(\ref{eq:2L}) and
(\ref{eq:alpha_approx}) lead to
\begin{equation}
  \label{eq:2Lapprox}
   \alpha  \approx \frac 14 \sum_{j=1}^{4}\alpha_j,
\end{equation}
where $\alpha$ is calculated for the whole, $2L\times2L$ system. Assuming that
$\alpha_j$ are independent random variables drawn from the same distribution
with mean 0, one arrives at
\begin{equation}
  \label{eq:sigma2L_L}
   \sigma_\alpha(2L)  \approx \frac{1}{2}\sigma_\alpha(L),
\end{equation}
which immediately leads to (\ref{eq:delta=1}).

Equation (\ref{eq:2Lapprox}) can be used iteratively to obtain
\begin{equation}
  \label{eq:2kL}
   \alpha(2^k L)  \approx \frac {1}{4^k} \sum_{j=1}^{4^k}\alpha_j(L), \quad k =
   1,2,\ldots
\end{equation}
where the arguments of $\alpha$ and $\alpha_j$ (i.e. $2^k L$ and $L$) indicate
the system size. The right-hand side of this formula is an arithmetic mean of
independent random variables with finite mean and variance, and---due to the central
limit theorem---converges to  normal distribution as $k\to\infty$. This explains
why the distribution of $\alpha$ for a sufficiently large system size $L$ can
be approximated by a normal distribution (see Fig.~\ref{fig:3}).

The above can be readily extended to flows in an arbitrary space dimension $d$.
We skip the details and report only the final conclusions. First,
\begin{equation}
 \label{eq:delta_d}
  \sigma_\alpha \propto L^{-\delta},\quad \delta = d/2
\end{equation}
for sufficiently large $L$. Second, the distribution of $\alpha$ tends to the
normal distribution as $L\to\infty$.

Equation (\ref{eq:delta_d}) implies that anisotropy effects  diminish with
system size most quickly in three-dimensional (3D) systems ($\sigma_\alpha
\propto L^{-3/2}$). Note, however, that the most important factor in computer
simulations is the total number of lattice nodes (or volume) $V$ in the system.
Using this quantity, equation (\ref{eq:delta_d}) can be written as
\begin{equation}
 \label{eq:sigmaV}
  \sigma_\alpha \propto V^{-1/2}
\end{equation}
irrespective of $d$. Thus, anisotropy of the permeability tensor should be
equally important (and difficult to account for) in computer simulations
carried out in any space dimension.

\begin{figure}
\includegraphics[width=0.9\columnwidth]{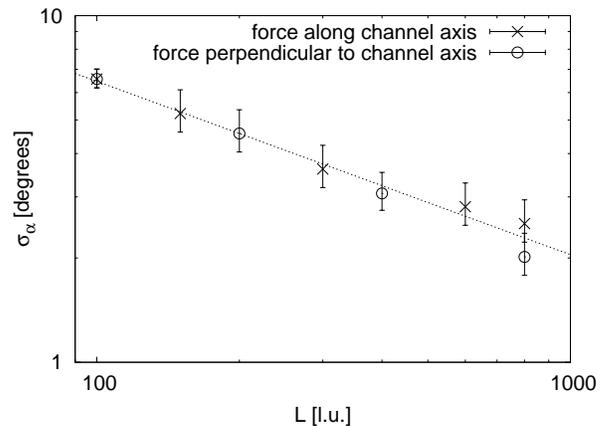}
\caption{$\sigma_\alpha$ for a quasi one-dimensional system of
 size $K\times L$ with $K$ fixed at $100$ l.u.\ and $L$ growing from $100$ to
 $800$ l.u.\ for channel axis parallel ($\times$) and perpendicular ({\large
 $\circ$})  to the external force ($a=4$ l.u., $\phi=0.7$, error bars at 95\%
 confidence level). Dashed line represents a single fit to $\sigma_\alpha
\propto 1/\sqrt{L}$ for all data points.
 \label{fig:9}}
\end{figure}

It is important to verify Eq.~(\ref{eq:delta_d}) for space dimensions $d \neq
2$. While at the moment our software cannot be used for $d=3$, the case $d=1$
can be tackled by studying a quasi one-dimensional system of size $K\times L$
with fixed $K$ and $L\to\infty$. Note that in this case Eq.~(\ref{eq:delta_d})
should hold irrespective of whether the longer side of the system is parallel
or perpendicular to the external force. The results, obtained for $a=4$ l.u.,
$\phi=0.7$, $K=100$, and $L$ ranging from $100$ to 800 l.u.\ are shown in
Fig.~\ref{fig:9} and confirm the validity of  Eq.~(\ref{eq:delta_d}).

%_______________________________ FIG-9
%
%
%
%_________________________________

Equations (\ref{eq:similarity}) and (\ref{eq:delta_d}) allow to factorize
$\sigma_\alpha(a,L,\phi)$:
\begin{equation}
 \label{eq:final}
     \sigma_\alpha = \left(\frac{a}{L}\right)^{d/2} \Phi(\phi),
\end{equation}
where $\Phi$ is a function. This relation can be expected to hold in general
only if $L \gg a$ and $\phi$ is sufficiently far away from the critical
porosity $\phi_c$. In a general case, $a$ is to be interpreted as a
characteristic system length (such as the diameter of discs, in case the
porous matrix is made of discs rather than squares), and $\Phi$ depends on the
system in question.

% ******************************************
% ************ Disscussion & Conclusions ***
% ******************************************

\section{Discussion and conclusions
\label{sec:discussion} }

Our results show that permeability anisotropy in statistically uniform porous
systems of sizes typically used in computer simulations is a significant
factor. The main parameter controlling this phenomena, especially at high
porosities, is the ratio $a/L$. For the model considered here, the asymptotic
regime is observed for $a/L \lesssim 0.04$. In this regime the distribution of
the angle $\alpha$ between the external force and the volumetric fluid flux is
very close to Gaussian, with the standard deviation diminishing as
$(a/L)^{d/2}$.

Although this conclusion is based on numerical results obtained for a
particular model of a two-dimensional flow, it is expected to apply to a wide
class of porous systems with randomly distributed identical solid matrices,
such as squares, disks or spheres. This observation can be used to estimate the
anisotropy-related statistical error in large-scale simulations, where often
only one large system is considered for each parameter set
\cite{Cancelliere90}. To this end it is enough to perform many independent
simulations in small- and medium-size systems, verify that
$\sigma_\alpha\propto (a/L)^{d/2}$, and extrapolate $\sigma_\alpha(L)$ to the
required value of $L$. Next, assuming that the distribution of $\alpha$ is
normal, one obtains the complete information about the error related to the
anisotropy of the permeability tensor.

Magnitude of permeability anisotropy could serve as a good indicator of how the
size of a model system compares with that of a REV.  We found that even for
$a/L=0.04$ the angle between the external force and the volumetric fluid flux
can be as large as $20^\circ$, and the permeability can vary with the
orientation of the  external force by a factor of 2. The value below which the
anisotropy effects are small enough to be practically negligible is $a/L
\approx 0.01 $, as in this case $\sigma_\alpha \lesssim 2^\circ$, i.e.\
$|\alpha| < 6^\circ$~with probability $p\approx0.99$. This enables to estimate
the size of a REV in the model considered here as $\approx 400\times400$ l.u.

It is interesting to note that most of the simulations carried out so far for
2D systems do not meet the criterion of $a/L \lesssim 0.01$, mainly because
they used models with large $a$.
In previous studies on two-dimensional flows in various statistically uniform
porous media, many researchers used $a/L$-values ranging from $0.02$
\cite{Andrade95}, through $0.026$ \cite{Verberg99}, 0.04
\cite{Kostek92,Martys92}, 0.05 \cite{Matyka08,Yiotis07} to 0.1
\cite{Koponen96,Koponen97,Nabovati07}, usually assuming their systems to be
isotropic. In view of our present findings, validity of this assumption in some
of these cases is questionable and requires verification. Generally, one should
expect that  the threshold value of $a/L$ below which the permeability
anisotropy is negligible is not universal, but depends on the geometry and
structure of the porous medium, especially on its porosity and space
dimensionality.

Anisotropy is a phenomenon independent of the boundary conditions. Periodic
boundary conditions used in this paper facilitate measurement of the
permeability tensor and reduce finite-size (boundary) effects. Other boundary
conditions could mask, but would not eliminate anisotropy effects. For example,
using solid walls along the fluid flow would fix the orientation of the fluid
stream, however, the system would respond to such boundaries with an internal
pressure gradient \cite{Bear72}, which would change (and complicate measurement
of) the orientation of the effective force acting on the fluid.

Finite-size permeability anisotropy in three-dimensional small porous systems
remains an open problem. Typical system sizes used in numerical 3D simulations
are $L\approx100$ l.u.\ The ratio $a/L$ is thus much larger in 3D than in 2D
simulations and ranges from $0.06$ \cite{Knackstedt94,Zhang95}, through $0.1$
\cite{Yiotis07}, $0.125$ \cite{Succi89,Cancelliere90}, to 0.33
\cite{Verberg99}. The magnitude of permeability anisotropy is usually
neglected. One exception is the paper by Verberg and Ladd \cite{Verberg99}, who
published the off-diagonal elements of the permeability tensor. Their data for
a single configuration of randomly distributed spheres suggests that
$\sigma_\alpha$ is a decreasing function of the porosity and varies from
$\sigma_\alpha \approx 3^\circ$ for $\phi = 0.464$ to $\sigma_\alpha\approx
18^\circ$ for $\phi=0.087$.  This is in agreement with our present findings for
a 2D system and indicates that permeability anisotropy is especially important
close to the percolation threshold.

\acknowledgments

We gratefully acknowledge support from UWr grant Nr~2944/W/IFT/08 (ZK, MM),
Sparda Bank M\"{u}nster eG (MM), and Max Planck Institute for Marine
Microbiology in Bremen (MM).

\end{document}